\documentclass[superscriptaddress,aps,prd,nofootinbib,twocolumn,showpacs]{revtex4-1}
\usepackage{amsfonts}
\usepackage{amsmath}
\usepackage{amssymb}
\usepackage{graphicx}
\usepackage{mathrsfs}
\usepackage{hhline,color}
\usepackage{pifont,ulem}
\usepackage{lmodern}
\usepackage{epsfig}
\usepackage{enumitem} 
\usepackage{ulem}
\usepackage[utf8]{inputenc}
\usepackage{multirow}

\newcommand{\pc}[1]{\ensuremath{\left(#1\right)}}

\newcommand{\ev}[1]{\ensuremath{\left\langle #1\right\rangle}}

\def\beq{\begin{equation}}
\def\eeq{\end{equation}}
\def\beqa{\begin{eqnarray}}
\def\eeqa{\end{eqnarray}}
\def\ban{\begin{eqnarray*}}
\def\ean{\end{eqnarray*}}
\def\bi{\begin{itemize}}
\def\ei{\end{itemize}}

\newcommand{\Z}{\mathbb{Z}}

\begin{document}

\title{Multiple critical endpoints in magnetized three flavor quark matter}

\author{M\'arcio Ferreira}
\email{mferreira@teor.fis.uc.pt}
\affiliation{CFisUC, Department of Physics,
University of Coimbra, P-3004 - 516  Coimbra, Portugal}
\author{Pedro Costa}
\email{pcosta@uc.pt}
\affiliation{CFisUC, Department of Physics,
University of Coimbra, P-3004 - 516  Coimbra, Portugal}
\author{Constan\c ca Provid\^encia}
\email{cp@fis.uc.pt}
\affiliation{CFisUC, Department of Physics,
University of Coimbra, P-3004 - 516  Coimbra, Portugal}

\date{\today}

\begin{abstract}
The magnetized phase diagram for three-flavor quark matter is studied within 
the Polyakov extended Nambu--Jona-Lasinio model.
The order parameters are analyzed with special emphasis on the strange quark
condensate.
We show that the presence of an external magnetic field induces several 
critical endpoints (CEPs) in the strange sector, which arise due to the 
multiple phase transitions that the strange quark undergoes.
The spinodal and binodal regions of the phase transitions are shown to increase 
with external magnetic field strength.
The influence of strong magnetic fields on the isentropic trajectories around 
the several CEPs is analyzed.
A focusing effect is observed on the region towards the CEPs that are related 
with the strange quark phase transitions.
Compared to the chiral transitions, the deconfinement transition turns 
out to be less sensitive to the external magnetic field and the crossover 
nature is preserved over the whole phase diagram.
\end{abstract}


\maketitle

\section{Introduction}

The possible existence of a critical endpoint (CEP) in the QCD phase diagram
is a long-standing issue that has captured the attention of the physics
community.
The nature of the phase transition between hadron matter and quark gluon 
plasma (QGP) can be inferred from its existence. 
A wide range of theoretical frameworks have been applied in analyzing the QCD
phase diagram and the possible existence of a CEP:
lattice QCD (LQCD) simulations \cite{Forcrand:2003hx,Fodor:2004nz}; 
Dyson-Schwinger equations \cite{Fischer:2014ata}; and several effective models, 
namely, the Nambu--Jona-Lasinio (NJL) model \cite{Costa:2008yh}, its extension 
up to eight-quark terms \cite{Moreira:2014qna}, and the 
Polyakov--Nambu--Jona-Lasinio (PNJL) model \cite{Costa:2008gr}.\\

The QGP expansion is recognized as a hydrodynamic expansion of an ideal 
fluid, which follows trajectories of constant entropy, known as isentropes. 
The conservation of the baryon number restricts the isentropic trajectories to
lines of constant entropy per baryon ($s/\rho_B$) in the ($T,\mu_B$) space
with zero strange quark density.
These isentropic trajectories contain important information on the adiabatic 
evolution of the system. 
For AGS, SPS, and RHIC the values of $s/\rho_B$ are 30, 45, and 300, 
respectively \cite{Bluhm:2007nu}. Lattice results for the (2+1)-flavor
equation of state (EOS) at these $s/\rho_B$ values are given in 
Refs. \cite{DeTar:2010xm,Borsanyi:2012cr}.
The presence of a CEP in the QCD phase diagram might deform the isentropes 
trajectories \cite{Asakawa:2008ti}.
This reinforces the importance of the search for the CEP, because 
modifications of the expansion trajectory may lead to observable effects in 
the hadron spectra (see Ref. \cite{Senger:2011zza}).\\

From the experimental point of view, the CEP existence/location is a
major goal of several heavy-ion-collisions (HIC) programs. 
The Beam Energy Scan (BES-I) program at the RHIC searched for experimental
signatures of the CEP, by colliding Au ions at several energies 
\cite{Abelev:2009bw}. 
Results of the moments of net-charge multiplicity distributions 
\cite{Adamczyk:2014fia} and of moments of net-kaon (proxy for net-strangeness)
multiplicity distributions \cite{Adamczyk:2017wsl} from the STAR Collaboration
provide relevant information on the freeze-out conditions (also for strange
quarks) and can help to clarify the existence of the CEP. 
However, high precision measurements and high statistics data are required for 
definitive conclusions regarding the CEP and precise determination of the
freeze-out conditions \cite{Adamczyk:2014fia,Adamczyk:2017wsl}.
High precision measurements for the net-kaon fluctuations will be made in the 
second phase of the RHIC BES.
Furthermore, QCD calculations should take into account the dynamics associated
with heavy-ion collisions before definitive conclusions about the CEP can be
made \cite{Adamczyk:2013dal}.
The results from the NA61/SHINE experiment at CERN SPS on the particle spectra 
and fluctuations (in p+p, Be+Be, and Ar+Sc collisions) show, so far, no 
indications of the existence of a CEP
\cite{Grebieszkow:2017gqx,Aduszkiewicz:2017mei}.
In the near future, planned experiments at FAIR (GSI) and NICA (JINR) will 
extend the search for a first-order phase transition and the CEP to regions of 
higher baryonic chemical potentials and lower temperatures, and hopefully 
unveil the existence and location of the CEP on the QCD phase diagram (see
\cite{Akiba:2015jwa} for a review on the experimental search of the CEP).\\

Strong external magnetic fields may play a role in multiple physical systems:
from HIC experiments at very high energies, to the early stages of the 
Universe, and astrophysical objects like magnetized neutron stars.
It becomes crucial, therefore, to understand how an external magnetic field
affects the structure of the QCD phase diagram.
Several LQCD calculations have been performed that address the impact of the 
magnetic field over the deconfinement and chiral transitions
\cite{baliJHEP2012,bali2012PRD,endrodi2013,Ilgenfritz:2013ara,D'Elia:2010nq,
Bornyakov:2013eya}. 
Besides the catalyzing effect of $B$ on dynamical chiral symmetry breaking, 
known as the magnetic catalysis (MC) effect (see \cite{Miransky:2015ava} for a 
review), LQCD results show also an additional effect: 
in the crossover transition region, the magnetic field, instead of catalyzing,
weakens the dynamical chiral symmetry breaking 
\cite{baliJHEP2012,bali2012PRD,endrodi2013}. 
This additional phenomenon is called the inverse magnetic catalysis (IMC) effect
(for a review see \cite{Andersen:2014xxa}).
Several low-energy effective models, including the Nambu--Jona-Lasinio 
(NJL)-type models, have been used to investigate the MC effect and its impact 
at finite chemical potential 
\cite{Fukushima:2012kc,Chao:2013qpa,Ferreira:2013tba,Ferreira:2013oda,
Farias:2014eca,Ferreira:2014exa,Ferreira:2014kpa,Ayala:2014iba,Ayala:2014gwa,
Ayala:2015lta,Ayala:2015bgv,Ayala:2016bbi,Pagura:2016pwr}.\\

The existence/location  of the CEP can be influenced by the medium strangeness 
and isospin content, and by the presence and strength of an external magnetic 
field \cite{Costa:2013zca}. 
Within the (2+1)-NJL model, it was verified that the CEP becomes located at 
higher temperatures with increasing $B$ \cite {Avancini:2012ee}. 
The same was obtained within the Ginzburg-Landau effective action formalism 
with the renormalized quark-meson model \cite{Ruggieri:2014bqa}. 
Using the (2+1)-PNJL model, the role played by vector interactions and the IMC 
effect on the CEP's location was analyzed in \cite{Costa:2015bza}, where 
opposite competing effects were found.
Another interesting aspect of the QCD phase diagram is the possible 
existence of a CEP associated with the strange quarks (with the respective 
first-order phase transition) in a generalized NJL model with the inclusion of 
explicit chiral symmetry breaking interactions \cite{Moreira:2014qna}. 
This implies the existence of two CEPs in the phase diagram. Indeed, the 
presence of these interactions acts as a catalyst in the production of strange 
quark matter when compared to conventional versions of the NJL model 
\cite{Moreira:2014qna}. 
Thus, it is interesting to explore in detail the effect of external magnetic 
fields on the strange sector, looking for the possible emergence of a CEP in 
this sector due to the catalyzing effect of $B$ at lower temperatures.\\

In the present work, we investigate the magnetized phase diagram using the 
(2+1)-flavor PNJL model. Special attention is devoted to the strange quark 
phase transition and the CEPs that appear in the presence of an external 
magnetic field. 
We investigate the isentropic trajectories near the light and strange CEPs, in 
both the crossover and first-order transition regions.
The model is presented in Sec. \ref{sec:model}, while the results are in Sec. 
\ref{sec:Results}.
Finally we draw our conclusions in Sec. \ref{Conclusions}.

\section{Model and Formalism}
\label{sec:model}
In the presence of an external magnetic field, the Lagrangian density of the 
PNJL model for (2+1)-flavor field takes the following form,
\begin{eqnarray}
{\cal L} &=& {\bar{q}} \left[i\gamma_\mu D^{\mu}-
	{\hat m}_c \right ] q ~+~ {\cal L}_\text{sym}~+~{\cal L}_\text{det}~-~\frac{1}{4}F_{\mu \nu}F^{\mu \nu}\nonumber\\
& &~+~ \mathcal{U}\left(\Phi,\bar\Phi;T\right) ,
	\label{Pnjl}
\end{eqnarray}
where the quark sector is described by the  SU(3) version of the NJL model 
\cite{Hatsuda:1994pi,Klevansky:1992qe}, which includes scalar-pseudoscalar, 
${\cal L}_\text{sym}$, and the 't Hooft six-fermion interaction, 
${\cal L}_\text{det}$, 
\begin{align}
	{\cal L}_\text{sym}&= G_s \sum_{a=0}^8 \left [({\bar q} \lambda_ a q)^2 + 
	({\bar q} i\gamma_5 \lambda_a q)^2 \right ] \\
	{\cal L}_\text{det}&=-K\left\{{\rm det} \left [{\bar q}(1+\gamma_5)q \right] + 
	{\rm det}\left [{\bar q}(1-\gamma_5)q\right] \right \}.
\end{align}

The quark field is represented in flavor space by
$q = (u,d,s)^T$, and ${\hat m}_c= {\rm diag}_f (m_u,m_d,m_s)$ 
is the corresponding (current) mass matrix. 
The Gell-Mann matrices are denoted by $\lambda_a$ ($0<a\le 8$) and 
$\lambda_0=\sqrt{2/3}I$, where $I$ represents the unit matrix. 
The coupling between the (electro)magnetic field $B$ and both the quarks and 
the effective gluon field is implemented  {\it via} the covariant derivative 
$D^{\mu}=\partial^\mu - i q_f A_{EM}^{\mu}-i A^\mu$, where $q_f$ represents the 
quark electric charge ($q_d = q_s = -q_u/2 = -e/3$); $A_{EM}^\mu$ is the 
external magnetic field 
($F^{\mu \nu }=\partial^{\mu }A_{EM}^{\nu }-\partial ^{\nu }A_{EM}^{\mu }$); 
and 
$A^\mu(x) = g_{strong} {\cal A}^\mu_a(x)\frac{\lambda_a}{2}$, where
${\cal A}^\mu_a$ is the SU$_c(3)$ gauge field.
We consider a  static and constant magnetic field in the $z$ direction, 
$A_{EM}^\mu=\delta^{\mu 2} x_1 B$.
The spatial components of the gluon field are neglected in the Polyakov gauge 
at finite temperature,
$A^\mu = \delta^{\mu}_{0}A^0 = - i \delta^{\mu}_{4}A^4$. 
The Polyakov loop is defined as the trace of the Polyakov line,
$ \Phi = \frac 1 {N_c} {\langle\langle \mathcal{P}\exp i\int_{0}^{\beta}d\tau\,
A_4\left(\vec{x},\tau\right)\ \rangle\rangle}_\beta$,
which is the order parameter of the $\Z_3$ 
symmetric/broken phase transition in pure gauge.\\

The pure gauge sector is described by the following effective potential 
$\mathcal{U}\left(\Phi,\bar\Phi;T\right)$ \cite{Roessner:2006xn},
\begin{eqnarray}
	& &\frac{\mathcal{U}\left(\Phi,\bar\Phi;T\right)}{T^4}
	= -\frac{a\left(T\right)}{2}\bar\Phi \Phi \nonumber\\
	& &
	+\, b(T)\mbox{ln}\left[1-6\bar\Phi \Phi+4(\bar\Phi^3+ \Phi^3)
	-3(\bar\Phi \Phi)^2\right],
	\label{Ueff}
\end{eqnarray}
where 
$a\left(T\right)=a_0+a_1\left(\frac{T_0}{T}\right)+a_2\left(\frac{T_0}{T}\right)^2$, 
$b(T)=b_3\left(\frac{T_0}{T}\right)^3$. 
The parameters were fitted to reproduce lattice results \cite{Roessner:2006xn}: 
$a_0 = 3.51$, $a_1 = -2.47$, $a_2 = 15.2$, 
and $b_3 = -1.75$.   
The critical temperature for the deconfinement phase transition is set by the 
parameter $T_0$, which in pure gauge was fixed to $T_0=270$ MeV. 

As a regularization scheme, we use a sharp cutoff $\Lambda$ in 
three-momentum space for the divergent ultraviolet sea quark integrals.  
For the model parametrization, we consider \cite{Rehberg:1995kh} 
$\Lambda = 602.3$ MeV, $m_u= m_d=5.5$ MeV, $m_s=140.7$ MeV, 
$G_s^0 \Lambda^2= 1.835$, and $K \Lambda^5=12.36$.\\

In the present study, we consider two model variants with distinct scalar 
couplings: constant $G_s=G_s^0$ and a magnetic field dependent $G_s=G_s(eB)$ 
\cite{Ferreira:2014kpa}.
In the latter case, the magnetic field dependence was determined 
phenomenologically, by reproducing the decrease in ratio of the chiral 
pseudocritical temperature obtained in LQCD calculations \cite{baliJHEP2012}.
Its functional dependence is 
$G_s(\zeta)=G_s^0\pc{\frac{1+a\,\zeta^2+b\,\zeta^3}{1+c\,\zeta^2+d\,\zeta^4}}$,
where $\zeta=eB/\Lambda_{QCD}^2$ (with $\Lambda_{QCD}=300$ MeV).
The parameters are $a = 0.0108805$, $b=-1.0133\times10^{-4}$, $c= 0.02228$, and 
$d=1.84558\times10^{-4}$ \cite{Ferreira:2014kpa}.
At zero magnetic field both models coincide, i.e., $G_s=G_s^0=G_s(eB=0)$.\\

For each value of temperature $T$, baryonic chemical potential $\mu_B$,
and magnetic field strength $B$, the mean field equations are obtained 
by minimizing the thermodynamic potential with respect to the order parameters 
\cite{Menezes:2008qt.Menezes:2009uc}: $\ev{u\bar{u}}$, $\ev{d\bar{d}}$, 
$\ev{s\bar{s}}$, $\Phi$, and $\bar{\Phi}$. 
Both the chiral and the deconfinement phase transitions can show different
natures: first-order, second-order, or crossover (analytic transition).
Contrarily to first-order phase transitions, the crossover transition is 
characterized by an analytic behavior, allowing for different definitions of 
(pseudo)critical temperature through different observables.
The pseudocritical temperature is often defined as the temperature at which 
the inflection point of the order parameters occurs.
Nevertheless, another possible definition is the temperature at which the 
order parameter reaches half its vacuum value, i.e., 
$\ev{q\bar{q}}(T,\mu_B)/\ev{q\bar{q}}(0,0)=0.5$ for quarks, 
and $\Phi (T,\mu_B)=0.5$ for the Polyakov loop. 
As we are going to analyze the order parameters via contour diagrams,
the latter definition of pseudocritical temperature will be useful.

\section{Results}
\label{sec:Results}

Herein, we consider the PNJL model with equal quark chemical 
potentials, $\mu_u=\mu_d=\mu_s=\mu_q$, which corresponds to zero charge 
(or isospin) chemical potential and zero strangeness chemical potential,
 i.e., $\mu_Q = \mu_S = 0$. The baryonic chemical potential is then given
by $\mu_B=3\mu_q$.

\subsection{Magnetized phase diagram}

To analyze how an external magnetic field affects the chiral/deconfinement 
transitions, we determine the quark condensates and the Polyakov loop value 
(order parameters) in the $(T,\mu_B)$ plane, for two magnetic field strengths: 
$eB=0.3$ GeV$^2$ and $eB=0.6$ GeV$^2$.
As we are mainly interested in examining where the phase transitions occur 
rather than on the specific condensates values, we normalize the condensates as
\begin{equation}
\ev{q\bar{q}}_0=\ev{q\bar{q}}_0(T,\mu_B,eB)=
 \frac{\ev{q\bar{q}}(T,\mu_B,eB)} {\ev{q\bar{q}}(0,0,eB)}.
\end{equation}
This way, regardless of the magnetic field strength, the normalized 
condensate $\ev{q\bar{q}}_0$ lies between 0 and 1. 
If one thinks about the quark masses instead, then we are looking at how the 
in-medium quark mass $M_q(T,\mu_B,eB)$ varies with respect to its magnetized 
vacuum value $M_q(0,0,eB)$.

\begin{figure}[!htbp]
	\centering
	\includegraphics[width=0.95\linewidth]{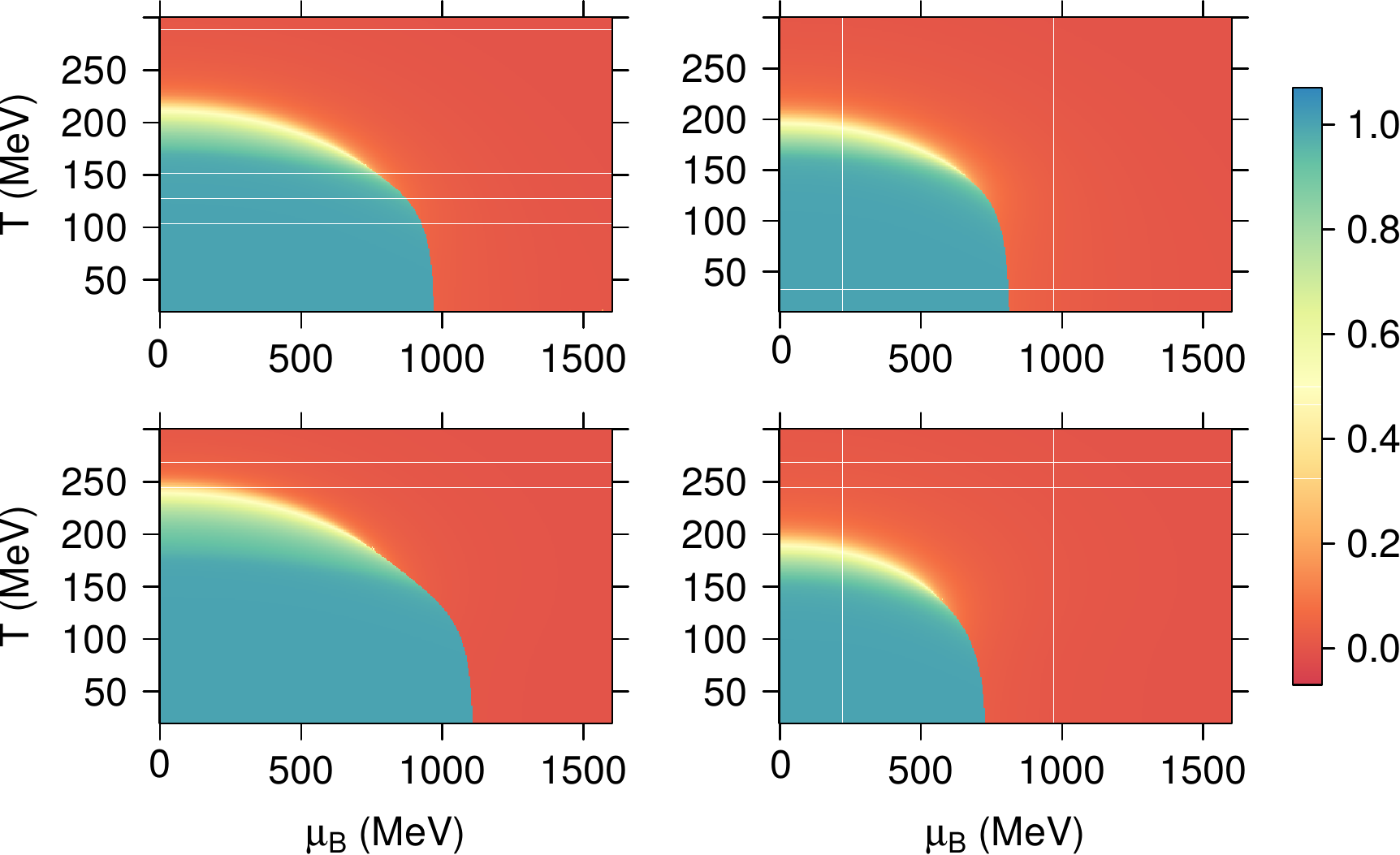}
	\caption{The normalized up-quark condensate $\ev{u\bar{u}}_0$
	(the color scale represents its magnitude) with $G_s^0$
		(left) and $G_s(eB)$ (right) for $eB=0.3$ GeV$^2$ (top) and $eB=0.6$ 
		GeV$^2$ (bottom).} 
	\label{fig:up_quark}
\end{figure}
The results for the normalized up-quark condensate $\ev{u\bar{u}}_0$ for both 
$G_s^0$ (constant coupling) and $G_s(eB)$ (magnetic coupling) models are in 
Fig. \ref{fig:up_quark}. 
The top panels show the results for $eB=0.3$ GeV$^2$, while the bottom panels 
have the $eB=0.6$ GeV$^2$ results. Furthermore, the left panels contain the 
$G_s^0$ model results, while the right panels display the $G_s(eB)$ model 
results.
The following conclusions can be drawn. 
At low temperatures, the chiral transition between the broken and the 
(approximately) restored regions is of first-order for all cases. 
The first-order transition occurs at $\mu^{crit}(T)$, at which the condensate 
changes abruptly from  $\ev{u\bar{u}}_0=1$ [or, looking at the quark mass, $M_u
(0,0,eB)$], to the (approximately) restored chiral symmetry with a much lower 
value of $\ev{u\bar{u}}_0$ ($M_u\approx m_u$).
Therefore, the chiral symmetry is restored via a strong (and unique)
first-order phase transition from the vacuum value to an almost zero value of 
$\ev{u\bar{u}}_0$.
The first-order transition persists in the phase diagram up to the CEP where 
the phase transition turns into second order.
Above the CEP's temperature, the transition shows an analytic nature 
(crossover transition).
The position of the CEP, $(T,\mu_B)_{\text{CEP}}$, is given in Table 
\ref{table:CEP_light} for each case.
\begin{table}[!b]
\begin{center}
    \begin{tabular}{|c||c|c|c||c|c|c|c|c|}
        \hline
				\multirow{3}{*}{CEP}	&\multicolumn{3}{c||}{$G_s=G_s^0$}&\multicolumn{3}{c|}{$G_s=G_s(eB)$} \\
				\cline{2-7}
        		& $T$ & $\mu_B$ & \multirow{2}{*}{$\rho_B/\rho_0$} & $T$ & $\mu_B$ & \multirow{2}{*}{$\rho_B/\rho_0$} \\
						& (MeV) & (MeV)	&  & (MeV)   & (MeV) &  \\
        \hline
        $eB=0$		& 157.5 & 890.4 & 1.74 & 157.5 & 890.4 	& 1.74	\\
        \hline
        $eB=0.3$  	& 192   & 674   & 3.54 & 177   & 627   	& 2.63	\\
        \hline
        $eB=0.6$ 	& 214 	& 692 	& 7.22 & 171 	& 535 		& 3.90  \\
        \hline
    \end{tabular}
    \caption{The temperature, baryonic chemical potential, and baryonic
		density (in units of $\rho_0=0.16$ fm$^{-3}$) at the light CEP, 
		$(T,\mu_B,\rho_B)_{\text{CEP}}$, for different values of $B$ (in GeV$^2$).}
    \label{table:CEP_light}
\end{center}
\end{table}
An important difference between the $G_s^0$ and $G_s(eB)$ models is clear: 
at zero temperature, the chemical potential at which the phase transition 
occurs, $\mu^{crit}_B(T=0)$, increases with $B$ for $G_s^0$, while the opposite 
happens for $G_s(eB)$.
Likewise, the pseudocritical temperature at $\mu_B=0$ decreases for $G_s(eB)$ 
]as expected, due to the $G_s(eB)$ parametrization] and increases for $G_s^0$.
Therefore, the overall effect of $G_s(eB)$ on the $\mu_B-T$ phase diagram is 
the diminishing of the region where chiral symmetry is broken.

The phase diagrams for the down-quark, $\ev{d\bar{d}}_0(T,\mu_B,eB)$, are 
similar to the up-quark results (Fig. \ref{fig:up_quark}) and are not shown. 
The main difference is that the crossover band, identified by the yellow band 
where $\ev{d\bar{d}}_0\approx0.5$, which defines the pseudocritical transition 
temperature, is located at slightly lower temperatures for the down quark, 
due to the electric charge difference.

\begin{figure}[!t]
	\centering
	\includegraphics[width=0.95\linewidth]{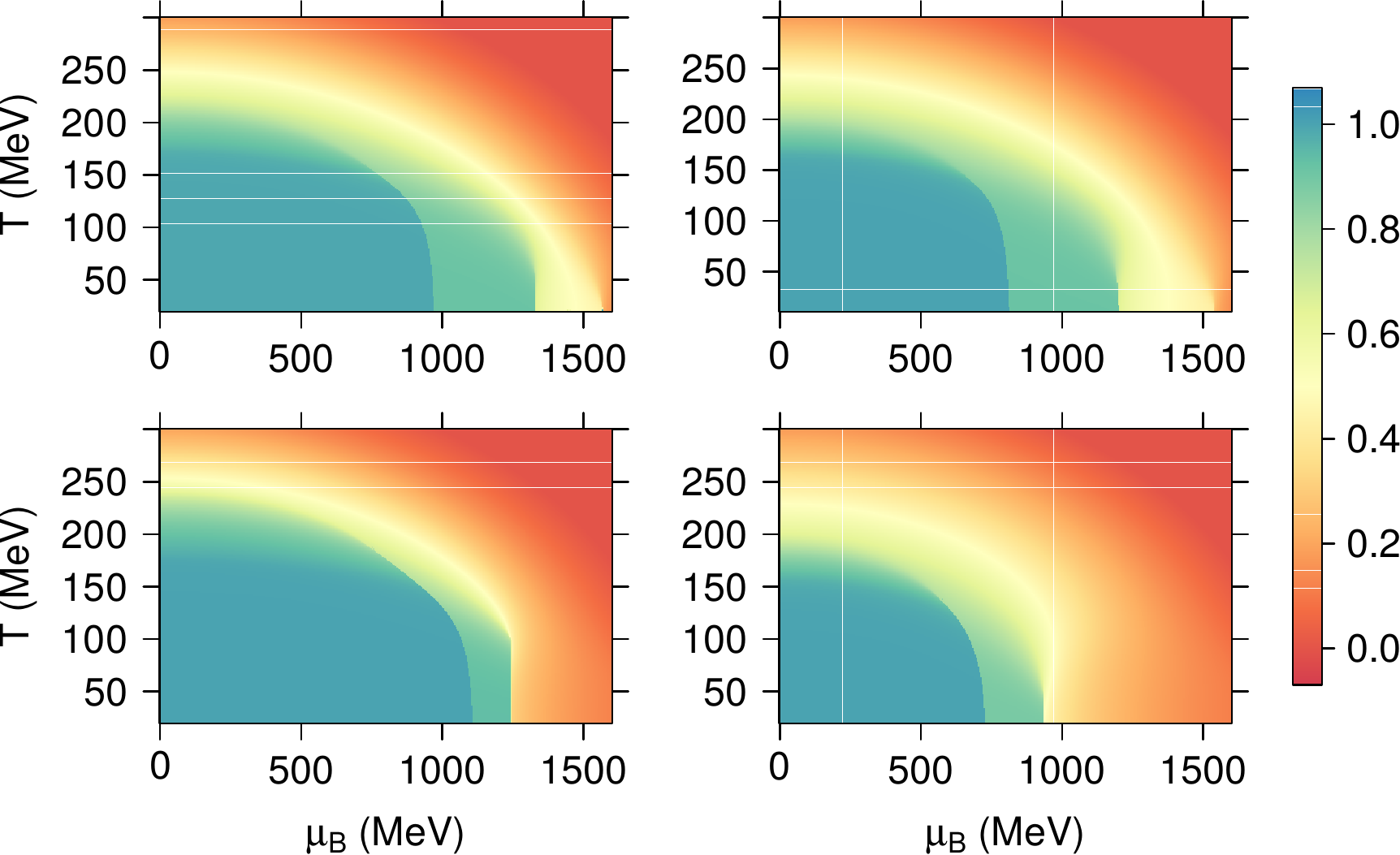}
	\caption{The normalized strange-quark condensate $\ev{s\bar{s}}_0$
	(the color scale represents its magnitude) with 
		$G_s^0$ (left) and $G_s(eB)$ (right) for $eB=0.3$ GeV$^2$ (top) and 
		$eB=0.6$ GeV$^2$ (bottom). } 
	\label{fig:strange_quark}
\end{figure}

The phase diagrams for the normalized strange-quark condensate 
$\ev{s\bar{s}}_0$  are displayed in Fig. \ref{fig:strange_quark} (same 
configuration as in Fig. \ref{fig:up_quark}), showing the following 
interesting features.
The condensate shows multiple discontinuities, indicating the presence of 
multiple first-order phase transitions, and thus the existence of multiple CEPs 
in the strange sector for all cases.

The existence of several first-order phase transitions at zero temperature
in the presence of an external magnetic field for the SU(2) NJL model was 
reported in \cite{Ebert:1999ht}.
Due to the Landau quantization induced by the magnetic field, instead of a 
single first-order transition, connecting the vacuum phase to the chirally 
restored phase, several intermediate first-order phase transitions take place. 
These complex patterns of multiple phase transitions were analyzed at zero 
temperature in \cite{Allen:2013lda,Denke:2013gha,Grunfeld:2014qfa},
where the number of first-order phase transitions, characterized by small 
jumps in the order parameters, were seen to grow as the magnetic field 
decreases.
At zero temperature, each phase transition can be attributed to the filling of 
a specific Landau level.
At finite temperature, even though all the Landau levels have a finite 
probability of being populated, at small temperatures, the multiple phase 
transitions can still be associated with the partial filling of the lower 
Landau levels. 
With increasing temperature, the number of phase transitions decreases to just 
one, and thus the multiple CEPs that appear in the phase diagram (one for each 
first-order phase transition) decreases to just one.
The several first-order phase transitions for the light sector, and the 
corresponding CEPs, were analyzed in \cite{Costa:2013zca}.
In Fig. \ref{fig:up_quark} just one first-order phase transition is present 
for the up-quark (and down-quark) because the magnetic fields considered are 
too high for multiple transitions to set in.
For smaller magnetic fields, several phase transitions, and corresponding CEPs,
are also present in the light sector (see \cite{Costa:2013zca}).

At $eB=0.3$ GeV$^2$ (top panels of Fig. \ref{fig:strange_quark}) both models 
show three first-order phase transitions that end up in three CEPs, while for 
$eB=0.6$ GeV$^2$ (lower panels of Fig. \ref{fig:strange_quark})
two first-order phase transitions are present.
Let $\mu^{crit}_i(T)$ denote the chemical potential at which the first ($i=1$), 
second ($i=2$), and third ($i=3$) first-order phase transitions take place, 
where $\mu^{crit}_1(T)<\mu^{crit}_2(T)<\mu^{crit}_3(T)$.
The first first-order phase transition, $\mu^{crit}_1(T)$, at which 
$\ev{s\bar{s}}_0$ has a small jump is induced on the strange quarks by the 
chiral transition of the light quarks (see Fig. \ref{fig:up_quark}). 
With the chiral symmetry already restored in the light sector for 
$\mu_B>\mu^{crit}_1(T)$, the two subsequent first-order phase transitions 
[i.e., $\mu^{crit}_2(T)$ and $\mu^{crit}_3(T)$], at which a sudden decrease of 
$\ev{s\bar{s}}_0$ occurs, can only be associated with the strange sector.
Therefore, at lower temperatures, the strange quarks undergo a phase 
transition from a region of broken chiral symmetry to an (approximately) 
restored one via intermediate transitions.
At $eB=0.3$ GeV$^2$, the strange quark takes the following values:
$\ev{s\bar{s}}_0$ for $\mu_B<\mu^{crit}_1(T)$, 
$0.9\ev{s\bar{s}}_0$ for $\mu^{crit}_1(T)<\mu_B<\mu^{crit}_2(T)$, 
$0.5\ev{s\bar{s}}_0$ for $\mu^{crit}_2(T)<\mu_B<\mu^{crit}_3(T)$, 
and $0.1\ev{s\bar{s}}_0$  for $\mu_B$ slightly above $\mu^{crit}_3(T)$ and,
then decreases smoothly with increasing $\mu_B$. 
The $G_s(eB)$ model predicts smaller values for $\mu^{crit}_1(T)$, 
$\mu^{crit}_2(T)$, and $\mu^{crit}_3(T)$ than the $G_s^0$ model.
At $eB=0.6$ GeV$^2$, the same pattern occurs but now with just two phase 
transitions. The location of the bright band, which indicates the value 
$(T,\mu_B)$ where $\ev{s\bar{s}}_0\approx0.5$, also shows that for $eB=0.3$, 
the first jump in the condensate at $\mu^{crit}_2(T)$ only reduces 
$\ev{s\bar{s}}_0$ slightly, and another second phase transition at 
$\mu^{crit}_3(T)$ is necessary to obtain a more complete restoration of the
chiral symmetry. 
The positions of the CEPs are listed in Table \ref{table:CEP_strange} for each 
case.

\begin{table}[!b]
\begin{center}
    \begin{tabular}{|c||c|c|c||c|c|c|c|c|}
        \hline
	\multirow{3}{*}{CEP}	&\multicolumn{3}{c||}{$G_s=G_s^0$}&\multicolumn{3}{c|}{$G_s=G_s(eB)$} \\
	\cline{2-7}
        & $T$ & $\mu_B$ & \multirow{2}{*}{$\rho_B/\rho_0$} & $T$ & $\mu_B$ & \multirow{2}{*}{$\rho_B/\rho_0$} \\
						& (MeV) & (MeV)	&  & (MeV)   & (MeV) &  \\
        \hline
        $eB=0$	& \multicolumn{3}{c||}{No CEP}& \multicolumn{3}{c|}{No CEP}							 \\
        \hline
        \multirow{2}{*}{$eB=0.3$}	& 62   & 1330  & 6.67 & 48   & 1193  & 5.40   \\
																		& 30   & 1566  & 11.10& 18   & 1539  & 10.75 \\
        \hline
        $eB=0.6$ 	& 124 & 1234& 12.10& 54  & 934 & 8.50   \\
        \hline
    \end{tabular}
    \caption{The temperature, baryonic chemical 
    potential, and baryonic density (in units of $\rho_0=0.16$ fm$^{-3}$)
    at the strange CEPs, $(T,\mu_B,\rho_B)_{\text{CEP}}$,
    for different values of $B$ (in GeV$^2$).}
		\label{table:CEP_strange}
\end{center}
\end{table}

\begin{figure}[!t]
	\centering
	\includegraphics[width=0.95\linewidth]{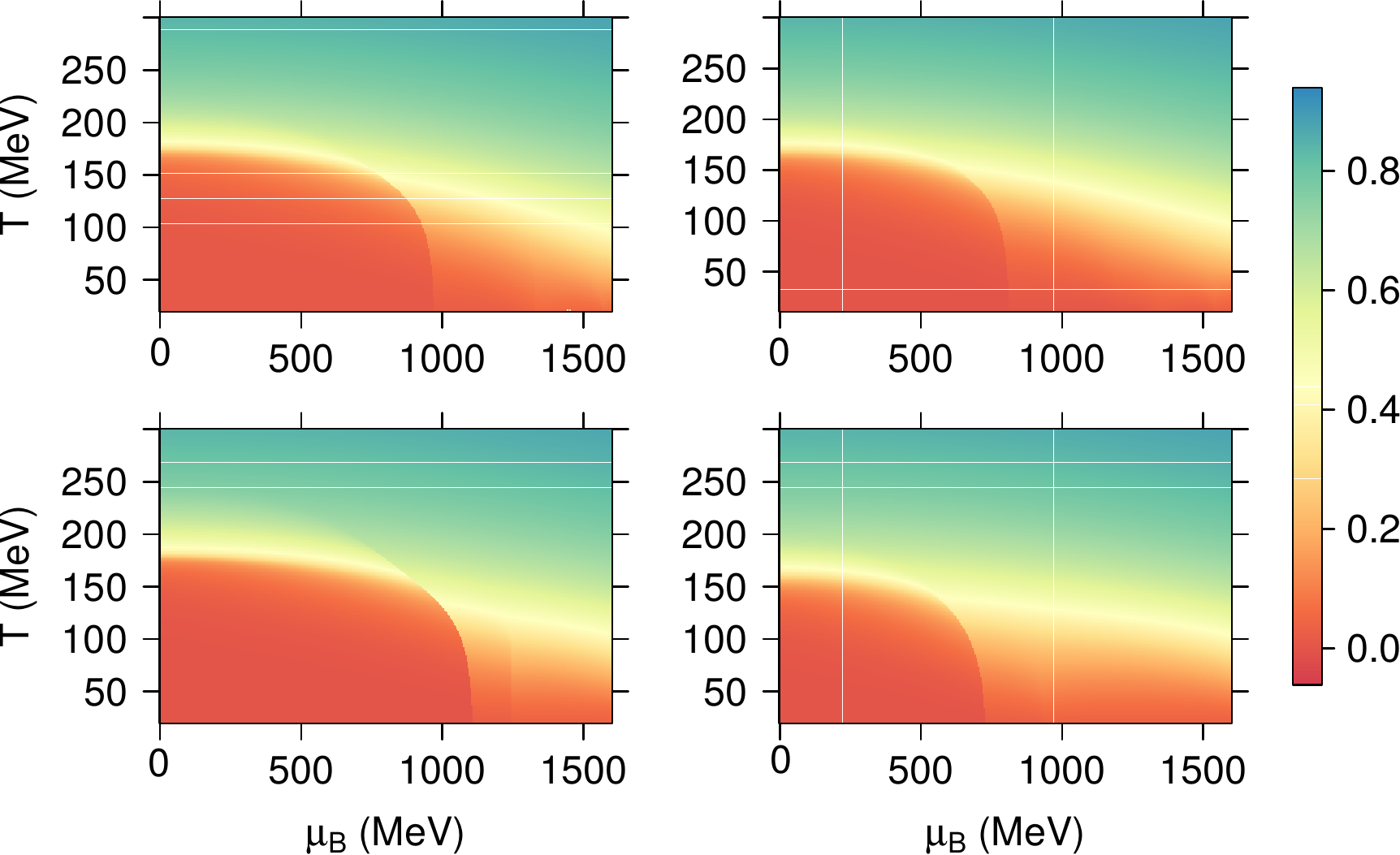}\\
	\caption{The Polyakov loop (the color scale represents its magnitude) with 
	$G_s^0$ (left) and $G_s(eB)$ (right) for $eB=0.3$ GeV$^2$ (top) and 
	$eB=0.6$ GeV$^2$ (bottom).} 
	\label{fig:Polyakov_loop}
\end{figure}

The phase diagram for the confinement/deconfinement transition, determined by 
the Polyakov loop value, $\Phi(T,\mu_B,eB)$, is presented in 
Fig.~\ref{fig:Polyakov_loop}.
The deconfinement pseudocritical temperature $T^{\Phi}$, defined by 
$\Phi(T^{\Phi},\mu_B)=0.5$, is a decreasing function of $\mu_B$ for all 
scenarios. 
The bright band represents a value of $\Phi\approx0.5$, and it thus can be used 
as a visual guide of the pseudocritical deconfinement transition in the 
$T-\mu_B$ plane. 
The discontinuity in the Polyakov loop value only reflects the first-order 
chiral phase transition for the light sector. For small values of $\mu_B$, 
the band of $\Phi\approx0.5$ is very close to the crossover region of the 
light sector. 
For low temperatures and high chemical potential values, there is a quark 
phase on which the chiral phase is already restored but confinement is 
still realized (low $\Phi$ values).
At $eB=0.6$ GeV$^2$, the $G_s^0$ model (bottom right panel) presents an 
intersection of the Polyakov loop line $\Phi=0.5$ with the first-order 
chiral phase transition line in the vicinity of the CEP for the light quarks. 
As the magnetic field increases, an overlapping occurs between the first-order 
phase transition, which moves to higher temperatures for the $G_s^0$ model, and 
the deconfinement transition, which in turn remains almost unchanged by the 
magnetic field presence. 

\subsection{Phase-separation boundaries}
\label{Phase-separation boundaries}

In this section we briefly analyze the quark phase transitions through the
phase-separation boundaries (binodals) and instability boundaries (spinodals). 

\begin{figure}[!htbp]
	\centering
	\includegraphics[width=1\linewidth,angle=0]{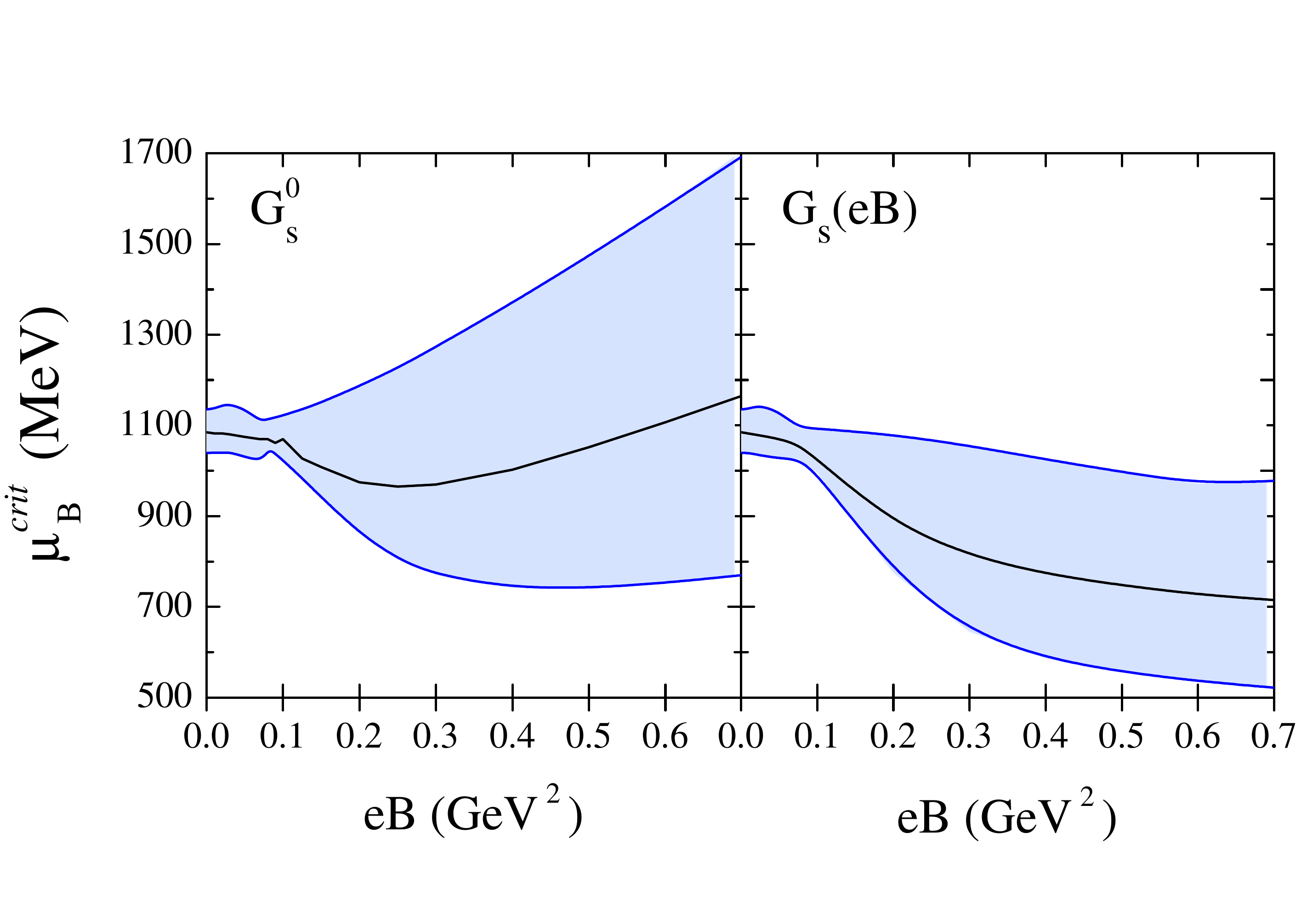}\\
	\caption{The spinodal (blue thick line) and 
	binodal (black thick line) $\mu_B^{crit}$ boundaries
	as a function of the magnetic field intensity at $T=0$.
	The models $G^0_s$ (left panel) and $G_s(eB)$ (right panel) are shown.} 
	\label{fig:mucrit}
\end{figure}

We first consider the phase-separation boundaries at zero temperature in
a $\mu_B-B$ plane. The results are in Fig. \ref{fig:mucrit}, where we display 
the spinodal lines (thick blue) and binodal line (thick black line) for the 
light quarks within both models. 
The spinodal region (blue area) increases with the magnetic field for both 
$G^0_s$ (left panel) and $G_s(eB)$ (right panel) models.

The pattern followed by the baryonic chemical potential value at which the 
light phase transition occurs at zero temperature, $\mu^{crit}_B(T= 0)$, was 
studied in detail for the PNJL model in \cite{Costa:2015bza}.
A lowering of $\mu^{crit}_B$ with $B$ was seen until $eB \approx 0.3$ GeV$^2$ 
for $G_s^0$, followed by a monotonically increasing of the 
$\mu_B^{\mbox{crit}}$ for stronger field strengths.
The $\mu^{crit}_B$ always decreases for the $G_s(eB)$ mode.
The existence of a range of magnetic fields, where at least two first-order 
phase transitions occur for the light sector, was also pointed out.
Therefore, instead of a single first-order transition, connecting the vacuum 
phase to the (approximately) chirally restored phase, we have multiple 
intermediate first-order phase transitions at $T=0$, that subsist at low 
and moderate temperatures, giving rise to multiple CEPs that appear in the 
phase diagram for the light sector \cite{Costa:2013zca,Costa:2015bza}. 
Indeed, the magnetic field induces a complex pattern of phase transitions for 
all quarks, i.e., for both the light (up and down) and the strange quarks.

Another clear feature also from Fig. \ref{fig:mucrit} is the spreading of the 
spinodal region in the $\mu_B$ direction with the increase of $B$ for both 
models, to a lower extent within the $G_s(eB)$ model. 
This has implications on the CEP location at finite temperatures, as seen in 
\cite{Costa:2015bza}. 
The temperature of the CEP is an increasing function of $B$ for the $G_s^0$ 
model, reflecting the increasing spreading of the spinodal region with $B$. 
The slower spreading of the spinodal region for the $G_s(eB)$ model, however, 
leads to an increase of the CEP's temperature only up to an intermediate 
$B$ strength. For higher $B$ fields, the width of the spinodal region remains 
approximately constant and the CEP's temperature remains almost unchanged
(as we will see in Fig. \ref{fig:CEP5}).

\begin{figure}[!t]
	\centering
	\includegraphics[width=0.92\linewidth,angle=0]{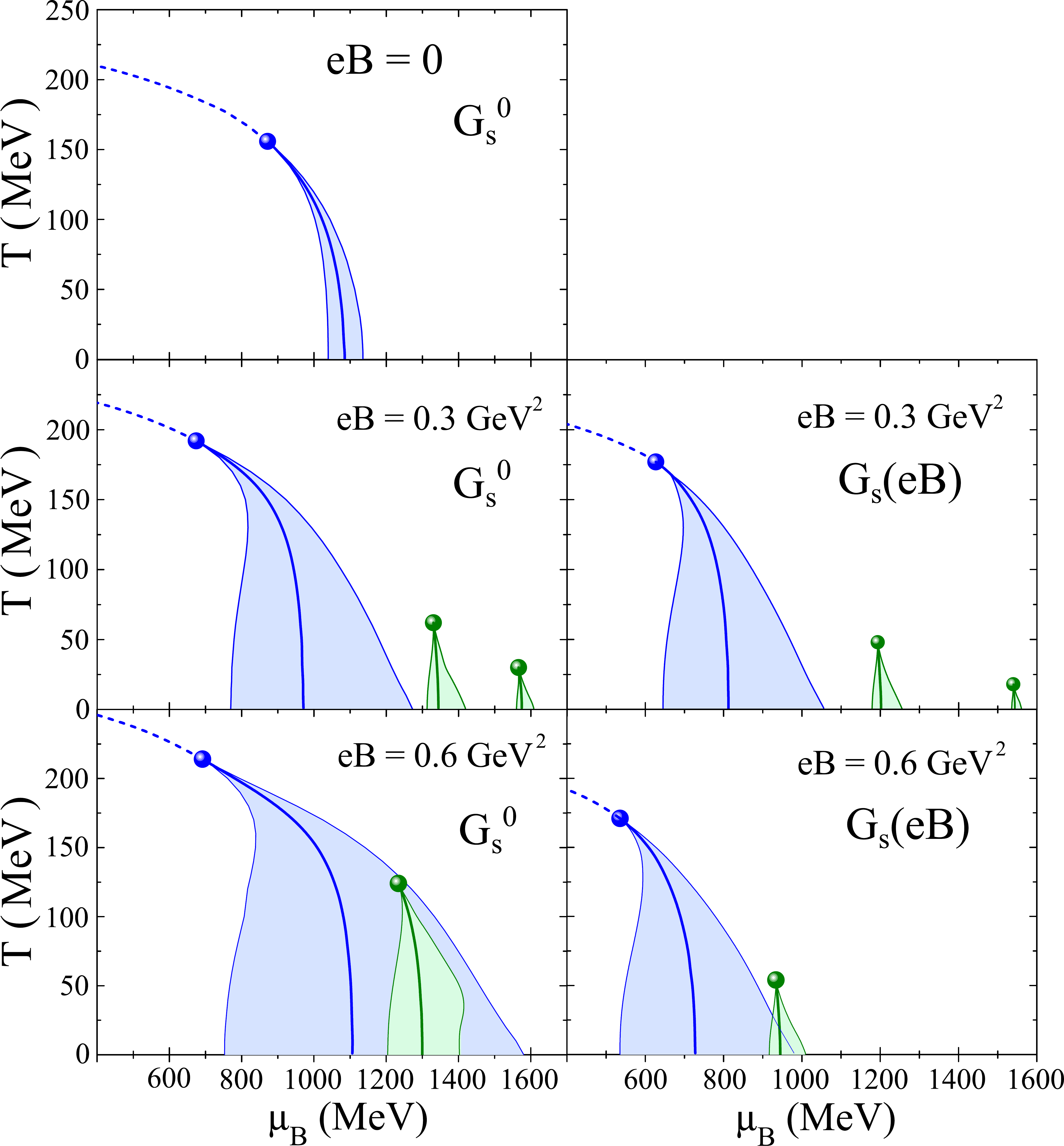}\\
	\caption{Binodal (thick lines) and spinodal (thin lines) regions in the
	temperature vs. baryonic chemical potential diagram for the light quarks 
	(blue) and strange quarks (green) at 	 $0$ (top), $0.3$ (middle), and 
	$0.6$ GeV$^2$. Both $G_s^0$ (left) and $G_s(eB)$ (right) model results are
	shown.} 
	\label{fig:CEP2}
\end{figure}

\begin{figure}[!htbp]
	\centering
	\includegraphics[width=0.92\linewidth,angle=0]{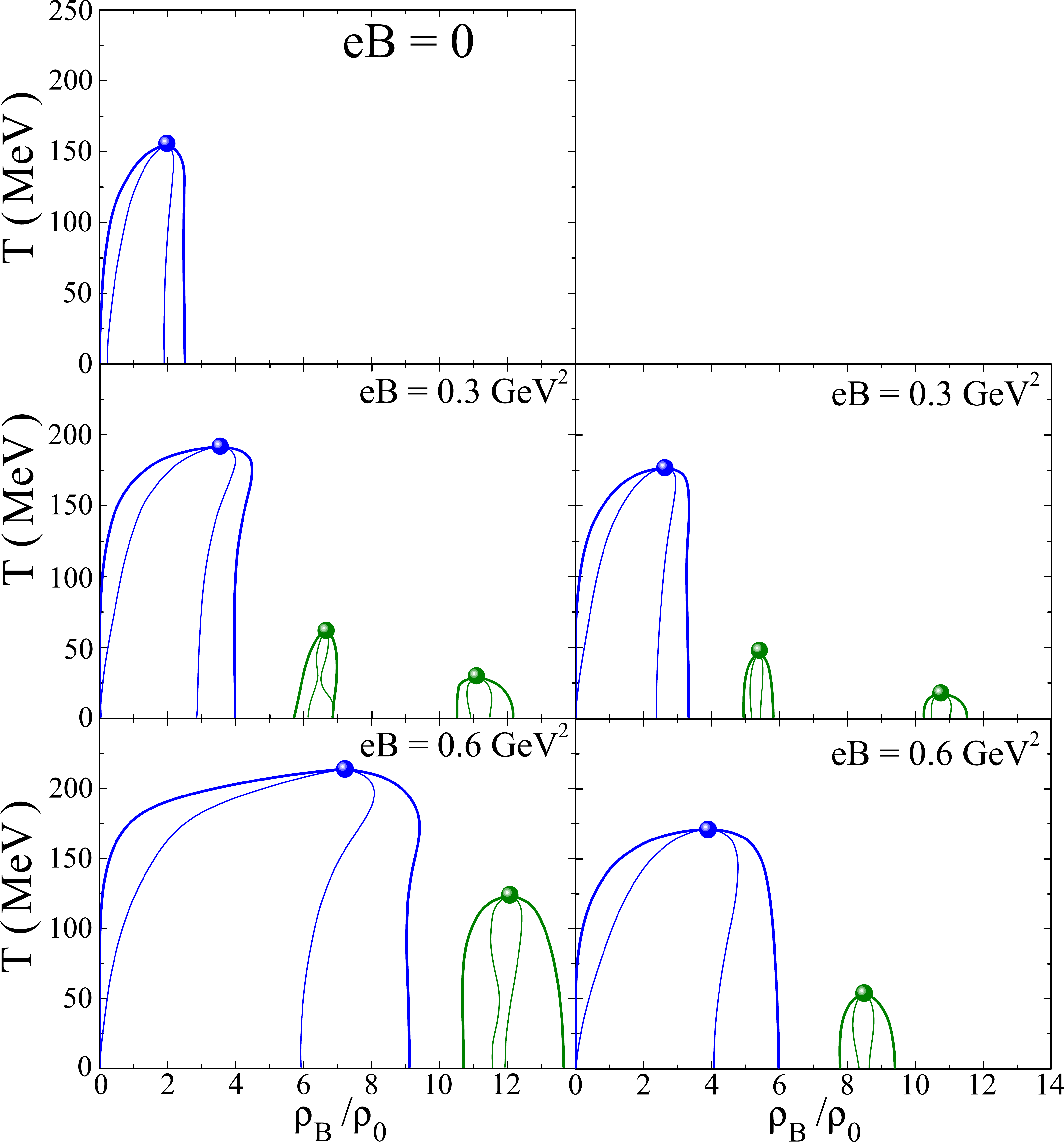}\\
	\caption{Binodal curve (thick lines) and spinodal section (blue region) in 
	the temperature vs. baryonic density diagram for the light quarks (blue) and 
	strange quarks (green) at $0$ (top), $0.3$ (middle), and $0.6$ GeV$^2$.
	Both $G_s^0$ (left) and $G_s(eB)$ (right) model results are shown
	 ($\rho_0=0.16$ fm$^{-3}$).
	 } 
	\label{fig:CEP3}
\end{figure}

In Figs. \ref{fig:CEP2} and \ref{fig:CEP3} we represent, respectively, $T-\mu_B$
and $T-\rho_B$ diagrams, where the binodals are represented by the thick lines 
and the spinodal by the thin lines.
The phase-separation boundaries are in blue for the light quarks and in green 
for the strange quark. 
Three magnetic field intensities are studied, $0$ (top), $0.3$ (middle), and 
$0.6$ GeV$^2$ (bottom), for the $G_s^0$ (right) and $G_s(eB)$ (left) models.
As already seen in Fig. \ref{fig:strange_quark}, we have two CEPs for the 
strange sector at $eB=0.3$ GeV$^2$. The existence of two CEPs for the strange 
sector occurs, for both $G_s^0$ and $G_s(eB)$ scenarios, in the range 
$0.2\gtrsim eB\gtrsim 0.4$ GeV$^2$. For $eB\lesssim 0.2$ GeV$^2$ more CEPs can 
exist due to the existence of numerous first-order transitions, while above 
0.4GeV$^2$ only one CEP persists.

Another important aspect is that, the stronger the magnetic field is, the 
larger the spinodal region becomes for both models, being the spinodal region
bigger when $G_s=G_s^0$. 
This is valid for both the light and strange transitions. The first-order 
lines are also shifted to lower values of $\mu_B$. For $eB=0.6$ GeV$^2$ (lower 
panels of Fig. \ref{fig:CEP2}) the spinodal regions in the ($T,\mu_B$) plane 
overlap with each other (blue for light quarks and green for strange quark). 
However, this happens at different baryonic densities (see lower panels of 
Fig. \ref{fig:CEP3}). The spinodal region for the strange quark is much 
smaller than for the light quarks and is located at higher baryonic densities.

From Fig. \ref{fig:CEP3} at zero temperature, we conclude that the upper 
baryonic densities at which the onset of both spinodal and binodal regions 
take place are increasing functions of $B$.

\subsection{The location of the critical endpoints}

In this section, we determine the location of the CEPs in the temperature vs. 
baryonic chemical potential diagram, and its dependence on the magnetic field 
strength.

\begin{figure}[!htbp]
	\centering
	\includegraphics[width=1.0\linewidth,angle=0]{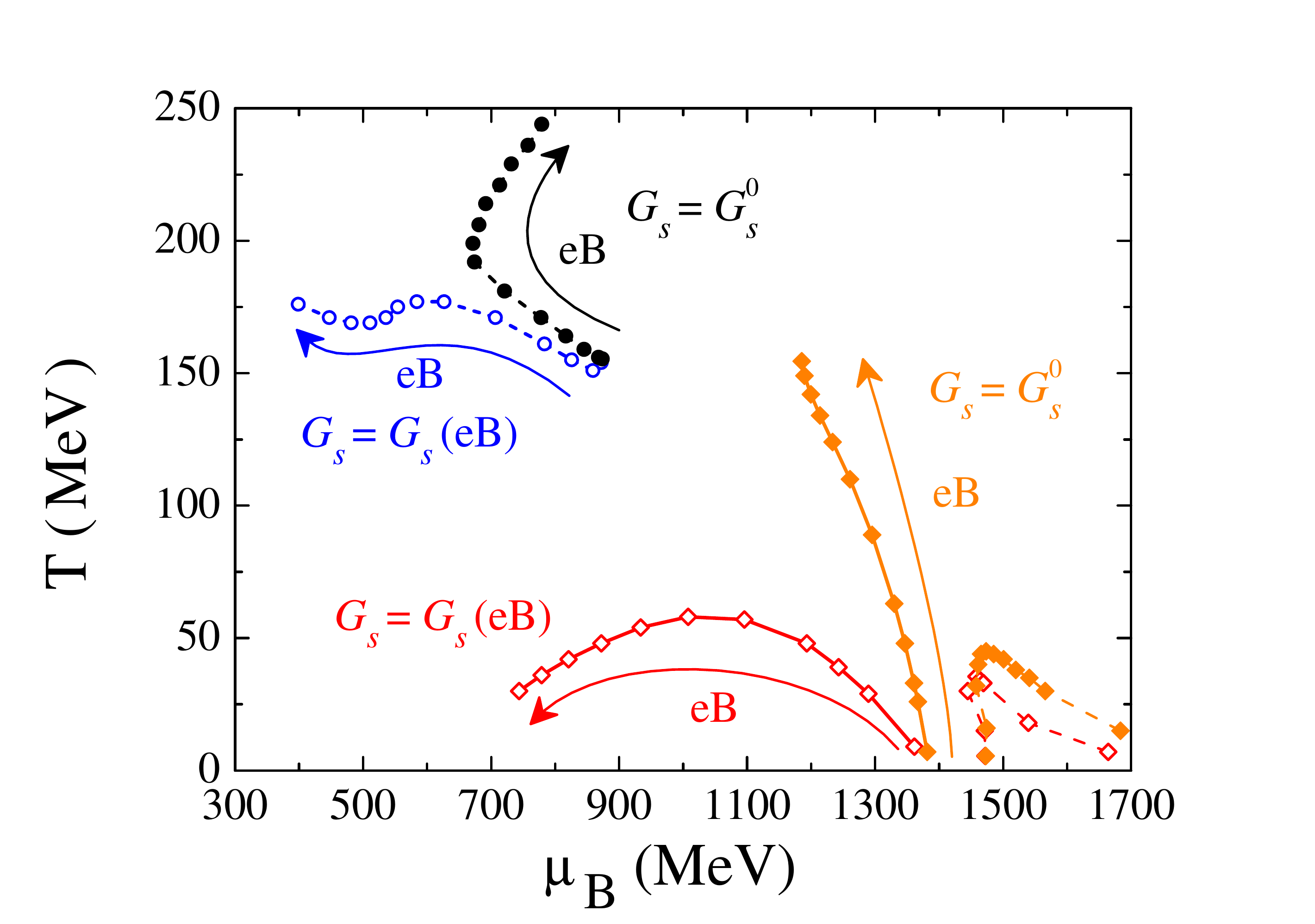}
	\caption{Critical endpoint of the light (blue and black) and strange 
	(red and orange) quarks as a function of the magnetic field intensity
	for the constant coupling, $G_s^0$, and magnetic dependent coupling, 
	$G_s(eB)$, models. The magnetic field increases from $0$ to $1$ GeV$^2$ in
	the arrows' directions.} 
	\label{fig:CEPs}
\end{figure}

Figure \ref{fig:CEPs} shows the location of the CEPs as a function of the 
magnetic field, $(T(B),\mu_B(B))_{\text{CEP}}$. 
The behavior of the CEP related with the chiral (light) transition (blue and 
black) was already reported in \cite{Costa:2015bza}. 
For moderate magnetic fields ($<0.3$ GeV$^3$) both models, $G_s^0$ and 
$G_s(eB)$, show similar results; i.e., the CEP moves towards higher 
temperatures and chemical potentials.
A distinctive behavior is seen for higher magnetic fields:
the CEP moves to lower $\mu_B$ for the $G_s(eB)$ model and the opposite occurs 
for the $G_s^0$ model. As already noticed in \cite{Costa:2015bza}, the 
$G_s(eB)$ model results indicate that, for high enough magnetic fields, the 
CEP moves towards the $\mu_B=0$ axis, and the analytic transition, present at 
$\mu_B=0$, will turn into a first-order phase transition.

Now let us focus on the CEP for the strange quark. 
As we had already seen in Fig. \ref{fig:strange_quark} (bottom right panel), 
the magnetic field induces multiple first-order phase transitions for the 
strange quark, and thus the existence of multiple CEPs.
In Fig. \ref{fig:CEPs} is shown two CEP branches for each model [red for 
$G_s(eB)$ and orange for $G_s^0$].
For both models, the CEP appearing at lower $\mu_B$ remains up to 
$eB\sim1$ GeV$^2$, while the CEP at higher $\mu_B$ disappears from the phase 
diagram at $eB\sim0.4$ GeV$^2$ (a similar behavior was already found for the 
light sector \cite{Costa:2013zca}).
The CEPs located at lower $\mu_B$ show a different behavior between models: 
while it moves towards lower $\mu_B$ in both models, at high magnetic 
fields $T$ increases monotonously with $B$ for $G_s^0$ and is a decreasing 
function for the $G_s(eB)$ model.
With increasing $B$, the CEP's location for the $G_s(eB)$ model (red) shows 
some similarity with the CEP of the light quarks (blue) by moving to lower 
$\mu_B$. 
For the $G_s^0$ model (orange) the CEP goes to lower values of $\mu_B$ but 
higher $T$.

\subsection{The isentropic trajectories}

\begin{figure*}[!htbp]
	\centering
	\includegraphics[width=0.9\linewidth,angle=0]{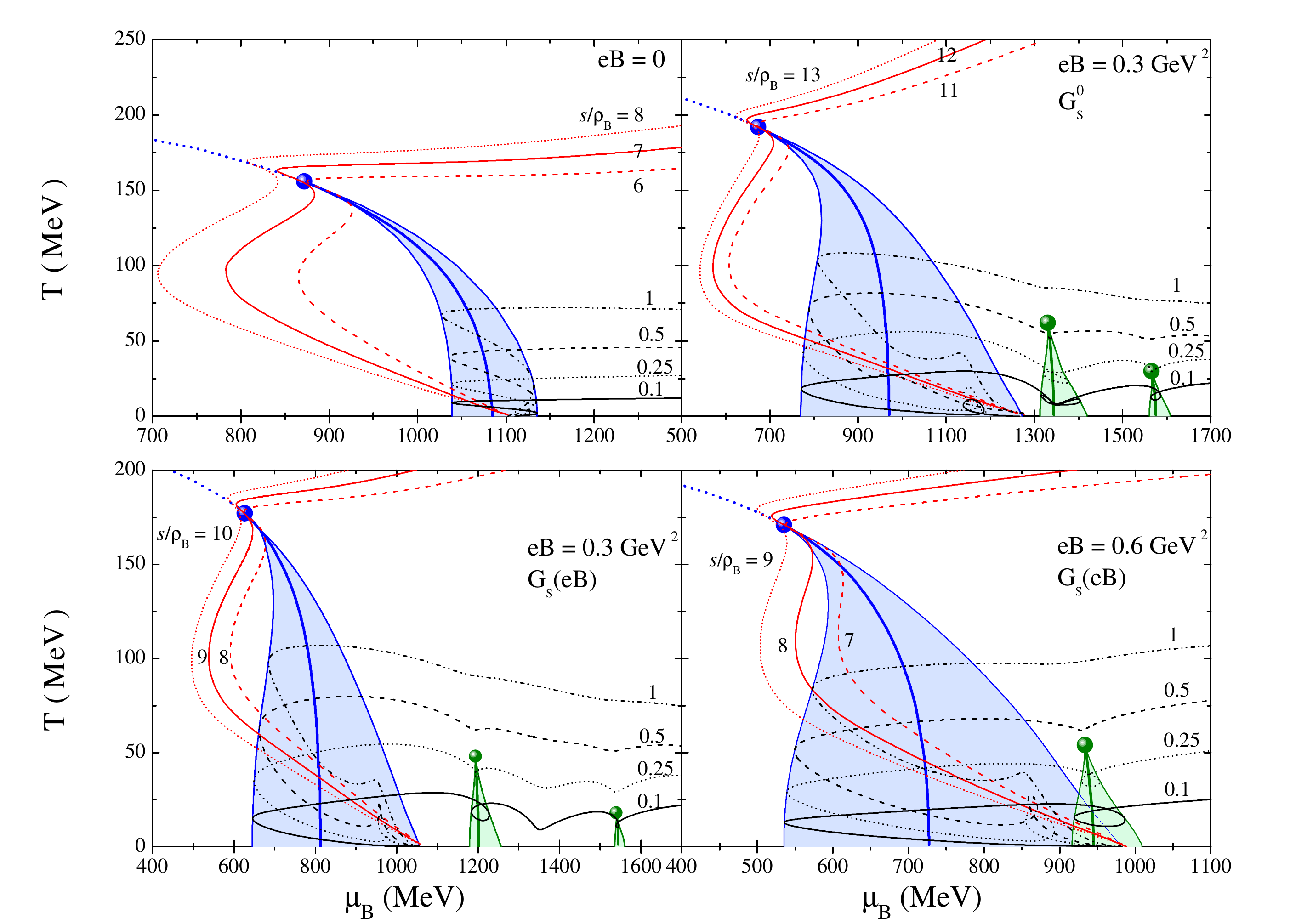}
	\caption{QCD phase diagram in the $T-\mu_q$ plane:
    the thick blue (green) lines represent the first-order phase transition 
		(binodal) while the thin blue (green) lines display the spinodal boundaries 
    for the light (strange) quarks.
    The isentropic trajectories for several values of $s/\rho_B$ are shown in 
		red and black.
    The following scenarios are considered:
    the $G_s^0=G_s(eB)$ model for $eB=0$ (top left panel),
    the $G_s^0$ model for $eB=0.3$ GeV$^2$ (top right panel),
    the $G_s(eB)$ model for $eB=0.3$ GeV$^2$ (bottom left panel),
    and the $G_s(eB)$ model for $eB=0.6$ GeV$^2$ (bottom right panel).
    The different scales in the $\mu_B$-axis allow us to clearly
		differentiate among the different isentropic trajectories.
    } 
	\label{fig:CEP4}
\end{figure*}
Finally, we analyze how the isentropic trajectories on the $T-\mu_B$ plane 
\cite{ Costa:2016vbb} and, in particular, near the CEPs are affected by the 
presence of an external magnetic field for both models, i.e., $G_s^0$ and 
$G_s(eB)$.
The interest in the isentropic trajectories relies on the hydrodynamical 
expansion of a HIC fireball that nearly follows trajectories of constant 
entropy.
New insights about the QCD phase diagram can thus be obtained by investigating 
these possible paths for the hydrodynamic evolution of a thermal medium 
created in the collisions and by studying the properties of matter
under these conditions.\\
  
We plot in Fig. \ref{fig:CEP4} several isentropic trajectories $s/\rho_B$
for both models and different magnetic field values on the $T-\mu_B$ plane. 
For the sake of comparability, the following scenarios have been selected 
in each panel: $eB=0$ (top left), $eB=0.3$ GeV$^2$ for the $G_s^0$ model (top 
right), $eB=0.3$ GeV$^2$ for the $G_s(eB)$ model (bottom left), and finally 
$eB=0.6$ GeV$^2$ for the $G_s(eB)$ model (bottom right).
For a clear distinction among the different isentropic trajectories, we have 
restricted the study to the range of $\mu_B$ values on which the binodals 
(metastable boundaries) occur in each case. 
From all possible isentropic lines (constant $s/\rho_B$) we have restricted 
ourselves to two sets: i) trajectories with higher values of $s/\rho_B$ 
represented by red lines that pass close to the CEP of the light quarks; ii) 
trajectories with $s/\rho_B\leq1$ displayed in black lines that go through the 
first-order phase transition line and, at larger $\mu_B$, pass near the CEPs 
connected with the strange sector.

In the following we discuss the behavior of the isentropic lines as 
temperature increases.
At zero temperature, all isentropic trajectories begin at the same $\mu_B$ 
value, i.e. $\mu_i=M_i$, which increases with $B$ (see Fig. \ref{fig:CEP4}). 
The temperature of the isentropic path $s/\rho_B$ takes a finite value 
as soon as $\rho_B$ becomes finite. 

For low values of the entropy per particle, a special pattern arises among 
the calculated isentropic trajectories (black lines): they are enclosed
within the spinodal boundary, which in the $T-\rho$ plane encloses
the unstable and metastable regions limited by the binodals.
Eventually, as the chemical potential further increases the isentropic line
leaves this region and proceeds towards the high $\mu_B$ chiral restored phase, 
keeping the temperatures approximately constant for $B=0$, but showing a 
decrease, or eventually, for the lower $s/\rho_B$ values, still 
an increase followed by a decrease for $eB=0.3,\, 0.6$ GeV$^2$. 
As we will discuss later, this is due to the onset of the strange quark, that
is pushed to lower values of $\mu_B$ and $\rho_B$ at finite $B$.

\begin{figure*}[!htbp]
	\centering
	\includegraphics[width=0.9\linewidth,angle=0]{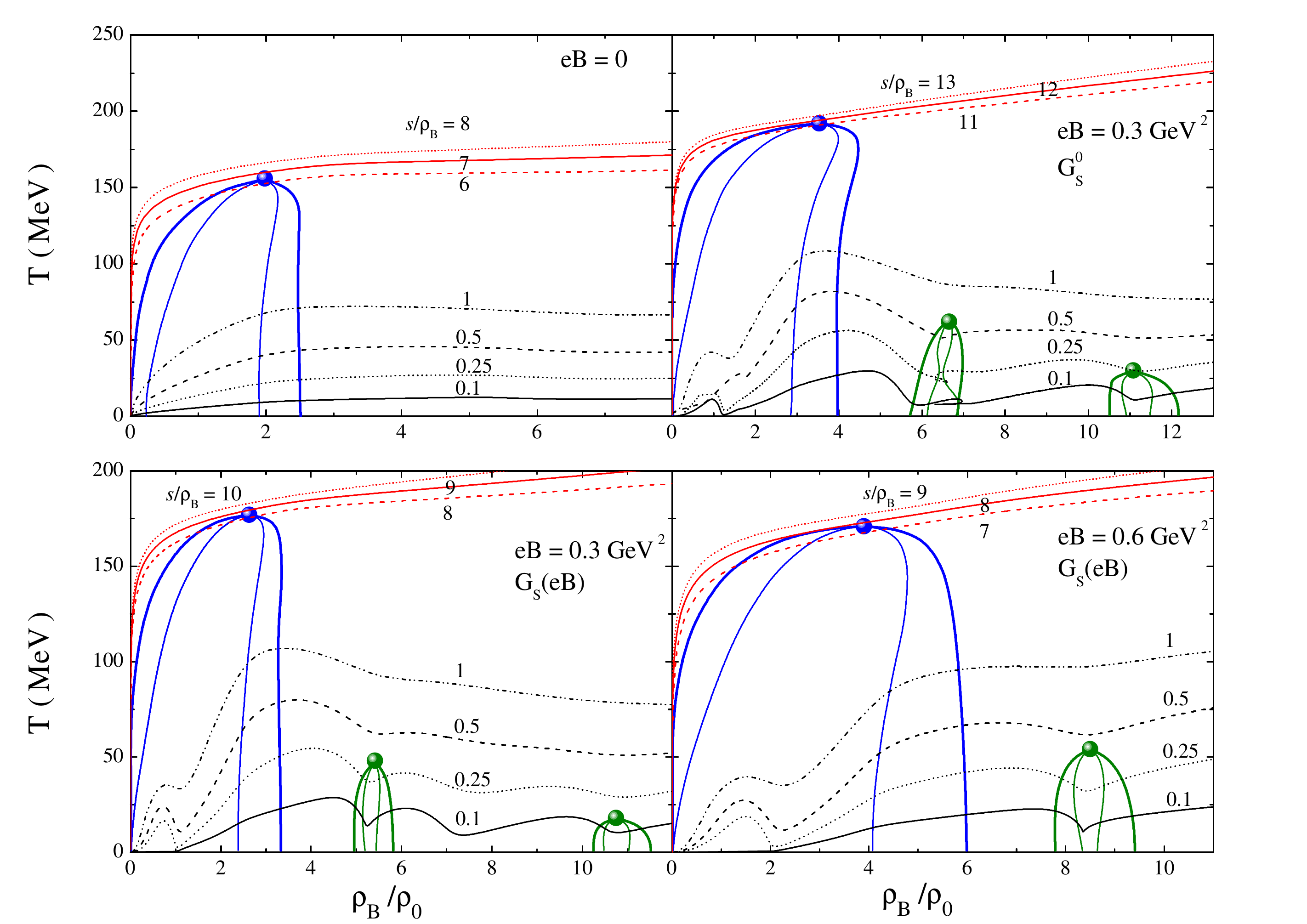}
	\caption{QCD phase diagram in the $T-\rho_B$ plane:
    the thick blue (green) lines represent binodal boundaries while the thin 
		blue (green) lines display the spinodal boundaries for the light (strange)
		quarks.
    The isentropic trajectories for several values of $s/\rho_B$ are shown in
		red and black.
    The following scenarios are considered:
    the $G_s^0=G_s(eB)$ model for $eB=0$ (top left panel),
    the $G_s^0$ model for $eB=0.3$ GeV$^2$ (top right panel),
    the $G_s(eB)$ model for $eB=0.3$ GeV$^2$ (bottom left panel),
    and the $G_s(eB)$ model for $eB=0.6$ GeV$^2$ (bottom right panel).
    The different scales in the $\rho_B$-axis allow us to clearly
		differentiate among the different isentropic trajectories.
		The baryonic density $\rho_B$ is represented in units of saturation
		density, $\rho_0=0.16$ fm$^3$.	
    } 
	\label{fig:CEP5}
\end{figure*}

For higher $s/\rho_B$ values (red lines), however, as the temperature 
increases from $T=0$ the trajectories cross the spinodal region entering the 
stable low density and chiral broken phase, moving towards the CEP, where
a kink occurs in the $T-\mu_B$ plane, and then move to higher $\mu_B$ (chiral 
restored phase) always with increasing temperature. 
As we will discuss later, this kink is not present in the $T-\rho_B$ plan (see 
Fig. \ref{fig:CEP5}) and is a feature of the presence of the CEP in the 
$T-\mu_B$ plane.

A different and interesting aspect is the trajectories' behavior near the 
strange quark CEPs, to which they are attracted.
Even though there is no focusing effect on the isentropic trajectories towards 
the CEP for the light quarks (see the red curves in Fig. \ref{fig:CEP4}), the 
CEPs related with the strange quark show a contrasting effect.
This behavior allows the prediction of other new CEPs if lower values of $B$ 
are analyzed: 
looking at the bottom left panel of Fig. \ref{fig:CEP4}, the bend present near 
$\mu_B=1350$ MeV signals that a CEP would emerge if we decrease the magnetic 
field strength.
The isentropes are quite affected by the growth of the spinodal region 
(related with the strengthening of the first-order transition due to the 
magnetic field), particularly for the light sector, and are pushed to higher 
$T$ in the transition region. The explanation for this behavior will be more 
clearly discussed looking to the phase diagram in a $T-\rho_B$ plane as will 
be done in the following.
  
\begin{figure*}[!htbp]
	\centering
	\includegraphics[width=0.9\linewidth,angle=0]{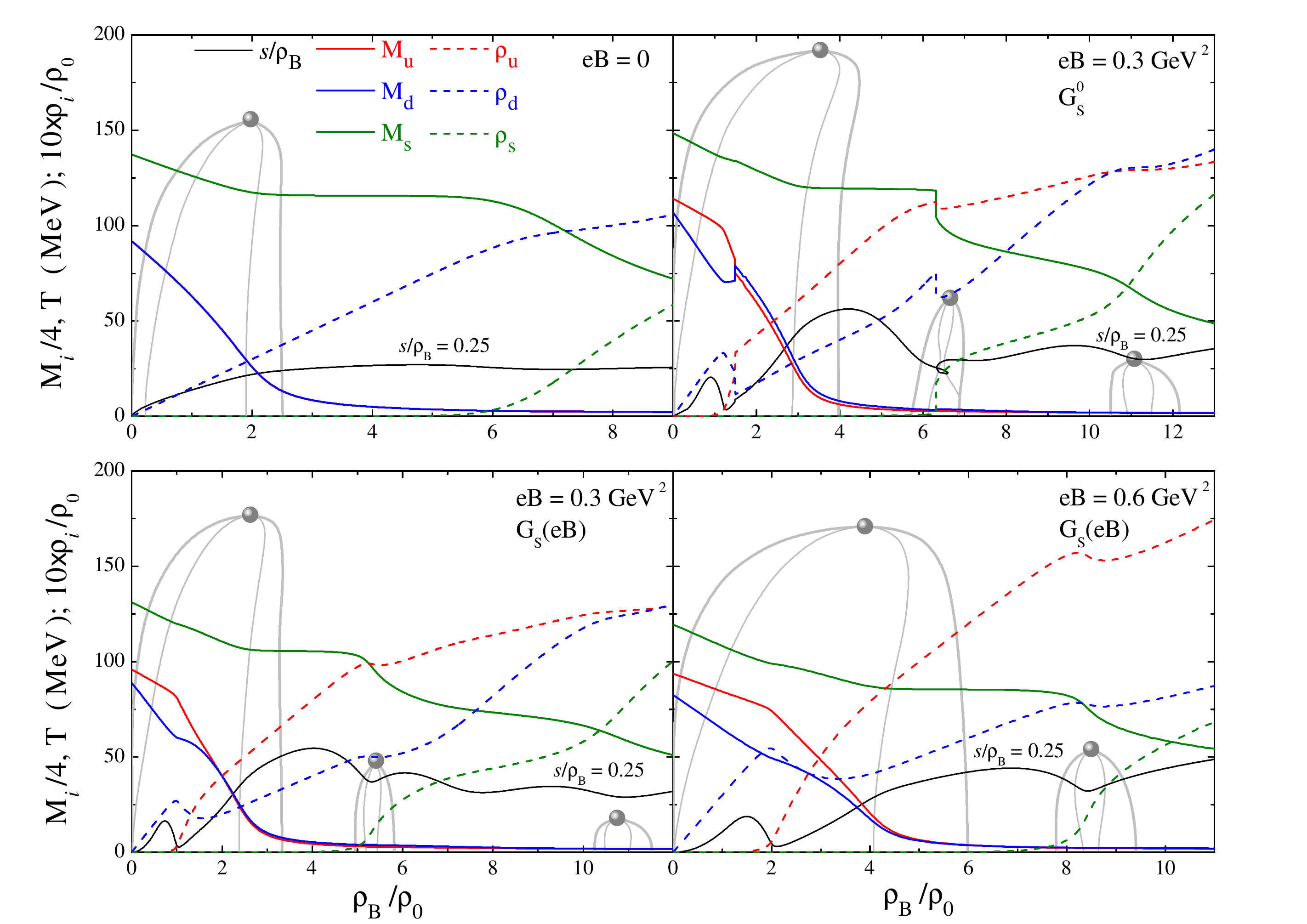}
	\caption{QCD phase diagram in the $T-\rho_B$ plane:
    the thick gray lines represent the binodal boundaries
    while the thin gray lines display the spinodal boundaries
    for the light and strange quarks.
    The isentropic trajectory for $s/\rho_B=0.25$ is shown as a black solid 
		line.
    The masses (solid lines) and densities (dashed lines) for each quark (up 
		in red, down in blue, and strange in green) are also displayed. 
    The following scenarios are considered:
    the $G_s^0=G_s(eB)$ model for $eB=0$ (top left panel),
    the $G_s^0$ model for $eB=0.3$ GeV$^2$ (top right panel),
    the $G_s(eB)$ model for $eB=0.3$ GeV$^2$ (bottom left panel),
    and the $G_s(eB)$ model for $eB=0.6$ GeV$^2$ (bottom right panel).
    The different scales in the $\rho_B$-axis allow us to clearly
		differentiate among the different isentropic trajectories.
		The baryonic density $\rho_B$ is represented in units of saturation
		density, $\rho_0=0.16$ fm$^3$.	
    } 
	\label{fig:isent025}
\end{figure*}

Finally, the shape of the isentropes also allows the perception of the 
spinodal region. Taking the line with $s/\rho_B=1$ for $eB=0$ at the lower 
temperatures (upper left panel) we see that this isentropic is bound by the
spinodal lines of the light sector. 
For finite $B$ the same effect is present. 
However, looking at Fig. \ref{fig:CEP4}, we find a loop structure for the 
$s/\rho_B=0.1$ and $s/\rho_B=0.25$ lines inside the spinodal region for the 
light sector (blue region). This is not related with the existence of a second 
first-order transition for the light sector but with the onset of the up quark 
(as we will see in Fig. \ref{fig:isent025}).

Additional insight can be attained by analyzing the $T-\rho_B$ diagrams. 
The results are shown in Figs. \ref{fig:CEP5} and \ref{fig:isent025} (with the same
configuration of Fig. \ref{fig:CEP4}). 
These phase diagrams show the density range of both unstable and metastable
regions.
In Fig. \ref{fig:CEP5}, the behavior of the trajectories with  higher $s/\rho_B
$ values (red lines) agrees for all scenarios: in the range of lower $\rho_B$ 
values, the required entropy per baryon is accomplished by a suddenly increase 
of temperature.
For $s/\rho_B\leq1$ (black lines) and $B\neq 0$, the isentropic trajectories 
show a nonmonotonic behavior in the $T-\rho_B$ plane.
This can be understood as follows:
the entropy abruptly increases when new degrees of freedom appear, such as a 
new quark species. 
Thus, to keep $s/\rho_B$ fixed, a suddenly decrease of temperature is needed 
to compensate this abrupt increase in entropy.
This can be seen in Fig. \ref{fig:isent025}, where we have plotted the 
isentropic line with $s/\rho_B=0.25$, the quark masses ($M_i$ with full 
colored lines), and quark densities ($\rho_i$ with dashed lines) for all 
scenarios. 
For instance, looking at the $G_s(eB)$ model results for $eB=0.6$ GeV$^2$ 
(right bottom panel), we conclude that in the range of densities 
$\rho_B \approx 1.5\rho_0-2.0\rho_0$ the temperature of the isentropic line 
decreases, and that the temperature also decreases at 
$\rho_B \approx 7.0\rho_0$. The reason is because at $\rho_B \approx 1.5\rho_0$
the density of the up quark becomes finite and at $\rho_B \approx 7.0\rho_0$ 
the strange quark density takes a nonzero value.  The same pattern is present 
for all scenarios.
Other less dramatic effects are related with the partial restoration of the 
chiral symmetry of the strange quark that occurs in several steps: a further 
decrease of $M_s$ gives rise to an increase of the strange quark density, and 
therefore a more equal distribution of $\rho_B$ among all quark flavors, but 
consequently a decrease of $T$ to keep $s/\rho_B$ constant.

The change of the properties of matter along the isentropes in the presence of 
a strong magnetic field will give rise to signatures of $B$ that could be 
identified. 
These could be
(a) a much higher abundance of $\pi^0$ pions at low densities
than the corresponding charged pions due to the late onset of the $u$-quarks or 
(b) the detection of a large amount of strange mesons. These features, however, 
require special matter conditions obtained from the  HIC, namely large 
densities and moderate temperatures.

\section{Conclusions}
\label{Conclusions}

We have studied the magnetized phase diagram for (2+1)-flavor quark matter 
within the PNJL model.
Besides the usual PNJL model with constant scalar coupling, we have also 
considered a magnetic field dependent coupling, which reproduces the IMC effect 
at $\mu_B=0$. 

The computed phase diagrams show that the nature of the deconfinement 
transition is quite insensitive to the external magnetic field strength for 
both models, preserving the analytic nature throughout the phase diagram.
The quark condensates show, however, a distinct behavior between models.
For the light quarks, the constant scalar coupling model gives rise to a 
region of broken chiral symmetry that increases with $B$, while for the 
magnetic field dependent coupling model it decreases with $B$.
The strange quark shows multiple first-order phase transitions at low 
temperatures, giving rise to multiple CEPs on the phase diagram. 
Therefore, the chiral symmetry on the strange sector is partially 
restored through multiple phase transitions.
The magnetic field induces a complex pattern of phase transitions not only for 
the strange quark, but also for the light quarks.
At higher temperatures, the strange quark undergoes an analytic transition
whose behavior and location are weakly model dependent.

We have analyzed the quark phase transitions through the phase-separation 
boundaries (binodals) and instability boundaries (spinodals).
For all flavors and within both models, the results show that the spinodal 
region grows with increasing $B$. However, the spinodal section associated 
with the strange quark is smaller and is located at higher baryonic densities.

We have studied how the multiple CEPs' locations vary when the magnetic field 
strength is increased.
Due to the occurrence of multiple first-order phase transitions, in both light 
and strange quarks, multiple CEPs emerge in the phase diagram.
For the strange quark, we have calculated the location of the two CEPs that 
appear at lower $\mu_B$ values. 
While the first CEP (at lower $\mu_B$) remains in the phase diagram up to 
$eB\sim1$ GeV$^2$, the second CEP (at higher $\mu_B$) disappears at 
$eB\sim0.4$ GeV$^2$. 
The location of the first CEP depends on the model: 
while, at lower $B$, it moves towards lower $\mu_B$ values in both models,
at higher $B$ it increases monotonically with $B$ for $G_s^0$ and decreases for 
$G_s(eB)$.

The isentropic trajectories in the $T-\mu_B$ and $T-\rho_B$ planes for both 
models and magnetic fields were calculated. 
The isentropes are affected by the growth of the spinodal region, particularly 
for the light sector, and are pushed to higher temperatures in the transition 
region, for large values of the entropy per baryon. 
Among the calculated values,   it was shown that the temperature along the 
isentropic lines for $s/\rho_B\leq1$ clearly indicates the appearance of new 
degrees of freedom or the (partial) restoration of chiral symmetry by  
decreasing with $\rho_B$ instead of increasing as at $B=0$. It is expected 
that the production of  mesons during the HIC reflects the composition of 
matter at a given density and, therefore, may act as signatures of the 
presence of an intense magnetic field.  
The CEPs related to the strange quark transitions show a focusing effect which 
is explained by the appearance of strangeness in matter. The focusing effect 
that occurs at larger densities was attributed to the partial restoration of the
chiral symmetry for the strange quark.

\vspace{0.25cm}
{\bf Acknowledgments.}
This work was supported by ``Fundação para a Ciência e Tecnologia'', Portugal, 
under the project No. UID/FIS/04564/2016, under the Grants No. 
SFRH/BPD/102273/2014 (P.C.), and under the project CENTRO-01-0145-FEDER-000014 
(M.F.) through CENTRO2020 program.
This work was partly supported by ``NewCompstar'', COST Action MP1304.


\begin{thebibliography}{99}

\bibitem{Fodor:2004nz} 
  Z.~Fodor and S.~D.~Katz,
  J. High Energy Phys. {\bf 0404}, 050 (2004),
  [hep-lat/0402006].
	
\bibitem{Forcrand:2003hx} 
  P.~de Forcrand and O.~Philipsen,
  Nucl.\ Phys.\ {\bf B673}, 170 (2003),
  [hep-lat/0307020].	

\bibitem{Fischer:2014ata} 
  C.~S.~Fischer, J.~Luecker, and C.~A.~Welzbacher,
  Phys.\ Rev.\ D {\bf 90}, no. 3, 034022 (2014),
  [arXiv:1405.4762 [hep-ph]].
	
\bibitem{Costa:2008yh} 
  P.~Costa, M.~C.~Ruivo, and C.~A.~de Sousa,
  Phys.\ Rev.\ D {\bf 77}, 096001 (2008),
  [arXiv:0801.3417 [hep-ph]];
  P.~Costa, C.~A.~de Sousa, M.~C.~Ruivo, and Y.~L.~Kalinovsky,
  Phys.\ Lett.\ B {\bf 647}, 431 (2007),
  [hep-ph/0701135].
    
\bibitem{Moreira:2014qna} 
  J.~Moreira, J.~Morais, B.~Hiller, A.~A.~Osipov, and A.~H.~Blin,
  Phys. Rev. D {\bf 91}, 116003 (2015),
  [arXiv:1409.0336 [hep-ph]].

\bibitem{Costa:2008gr} 
  P.~Costa, C.~A.~de Sousa, M.~C.~Ruivo, and H.~Hansen,
  Europhys.\ Lett.\  {\bf 86}, 31001 (2009),
  [arXiv:0801.3616 [hep-ph]];
  P.~Costa, M.~C.~Ruivo, C.~A.~de Sousa, and H.~Hansen,
  Symmetry {\bf 2}, 1338 (2010),
  [arXiv:1007.1380 [hep-ph]].	
	
\bibitem{Bluhm:2007nu} 
  M.~Bluhm, B.~Kampfer, R.~Schulze, D.~Seipt and U.~Heinz,
  Phys.\ Rev.\ C {\bf 76}, 034901 (2007),
  [arXiv:0705.0397 [hep-ph]].
    
\bibitem{DeTar:2010xm} 
  C.~DeTar, L.~Levkova, S.~Gottlieb, U.~M.~Heller, J.~E.~Hetrick, R.~Sugar and D.~Toussaint,
  Phys.\ Rev.\ D {\bf 81}, 114504 (2010),
  [arXiv:1003.5682 [hep-lat]].

\bibitem{Borsanyi:2012cr} 
  S.~Borsanyi, G.~Endrodi, Z.~Fodor, S.~D.~Katz, S.~Krieg, C.~Ratti and K.~K.~Szabo,
  J. High Energy Phys. {\bf 1208} (2012) 053,
  [arXiv:1204.6710 [hep-lat]].

\bibitem{Asakawa:2008ti} 
  M.~Asakawa, S.~A.~Bass, B.~Muller and C.~Nonaka,
  Phys.\ Rev.\ Lett.\  {\bf 101}, 122302 (2008),
  [arXiv:0803.2449 [nucl-th]].

\bibitem{Senger:2011zza}
  P. Senger, E. Bratkovskaya, A. Andronic, R. Averbeck, R. Bellwied, {et~al.},
  {Lect.Notes Phys.} {\bf 814} (2011) 681--847.

\bibitem{Abelev:2009bw} 
  B.~I.~Abelev {\it et al.}  (STAR Collaboration),
  Phys.\ Rev.\ C {\bf 81}, 024911 (2010),
  [arXiv:0909.4131 [nucl-ex]].

\bibitem{Adamczyk:2014fia} 
  L.~Adamczyk {\it et al.}  (STAR Collaboration),
  Phys.\ Rev.\ Lett.\  {\bf 113}, 092301 (2014),
  [arXiv:1402.1558 [nucl-ex]].

\bibitem{Adamczyk:2017wsl}
  L.~Adamczyk {\it et al.} (STAR Collaboration),
  arXiv:1709.00773 [nucl-ex].

\bibitem{Adamczyk:2013dal} 
  L.~Adamczyk {\it et al.}  (STAR Collaboration),
  Phys.\ Rev.\ Lett.\  {\bf 112}, no. 3, 032302 (2014),
  [arXiv:1309.5681 [nucl-ex]].

\bibitem{Grebieszkow:2017gqx} 
  K.~Grebieszkow [NA61/SHINE Collaboration],
  arXiv:1709.10397 [nucl-ex].

\bibitem{Aduszkiewicz:2017mei} 
  A.~Aduszkiewicz [NA61/SHINE Collaboration],
  Nucl.\ Phys.\ A {\bf 967}, 35 (2017),
  [arXiv:1704.08071 [hep-ex]].

\bibitem{Akiba:2015jwa} 
  Y.~Akiba {\it et al.},
  arXiv:1502.02730 [nucl-ex].

\bibitem{endrodi2013} 
  F.~Bruckmann, G.~Endr\"odi, and T.~G.~Kovacs,
  J. High Energy Phys. {\bf 1304}, 112 (2013),
  [arXiv:1303.3972 [hep-lat]].

\bibitem{baliJHEP2012}
	G. S. Bali, F. Bruckmann, G. Endr\"odi, Z. Fodor, S. D. Katz, S. Krieg, 
  A. Sch\"afer, and K. K. Szab\'o,
  J. High Energy Phys. {\bf 1202}, 044 (2012),
  [arXiv:1111.4956 [hep-lat]].

\bibitem{bali2012PRD}
	G. S. Bali, F. Bruckmann, G. Endr\"odi, Z. Fodor, S. D. Katz, and A. Sch\"afer,
  Phys.\ Rev.\ D {\bf 86}, 071502 (2012),
  [arXiv:1206.4205 [hep-lat]].

\bibitem{Ilgenfritz:2013ara}
E.-M.~Ilgenfritz, M.~Muller-Preussker, B.~Petersson, and A.~Schreiber,
  Phys.\ Rev.\ D {\bf 89}, 054512 (2014),
  arXiv:1310.7876 [hep-lat].
  
\bibitem{D'Elia:2010nq}
  M.~D'Elia, S.~Mukherjee, and F.~Sanfilippo,
  Phys.\ Rev.\ D {\bf 82}, 051501 (2010),
  [arXiv:1005.5365 [hep-lat]].
  
\bibitem{Bornyakov:2013eya} 
  V.~G.~Bornyakov, P.~V.~Buividovich, N.~Cundy, O.~A.~Kochetkov, and A.~Sch\"afer,
	Phys. Rev. D {\bf 90}, 034501 (2014),
  arXiv:1312.5628 [hep-lat].
 
\bibitem{Miransky:2015ava} 
V.~A.~Miransky and I.~A.~Shovkovy,
Phys.\ Rep.\  {\bf 576}, 1 (2015),
[arXiv:1503.00732 [hep-ph]].


\bibitem{Andersen:2014xxa} 
J.~O.~Andersen, W.~R.~Naylor and A.~Tranberg,
Rev.\ Mod.\ Phys.\  {\bf 88}, 025001 (2016),
[arXiv:1411.7176 [hep-ph]].

\bibitem{Fukushima:2012kc} 
  K.~Fukushima and Y.~Hidaka,
  Phys.\ Rev.\ Lett.\  {\bf 110}, no. 3, 031601 (2013)
  [arXiv:1209.1319 [hep-ph]].
 
\bibitem{Chao:2013qpa} 
  J.~Chao, P.~Chu, and M.~Huang,
  Phys.\ Rev.\ D {\bf 88}, 054009 (2013)
  [arXiv:1305.1100 [hep-ph]].  

\bibitem{Ferreira:2013tba} 
  M. Ferreira, P. Costa, D.P. Menezes, C. Provid\^{e}ncia, and N.N.~Scoccola,
  Phys.\ Rev.\ D {\bf 89}, 016002 (2014),
  [arXiv:1305.4751 [hep-ph]].

\bibitem{Ferreira:2013oda} 
  M.~Ferreira, P.~Costa, and C.~Providência,
  Phys.\ Rev.\ D {\bf 89}, no. 3, 036006 (2014),
  [arXiv:1312.6733 [hep-ph]].
  
\bibitem{Ferreira:2014kpa} 
  M.~Ferreira, P.~Costa, O.~Lourenço, T.~Frederico, and C.~Providência,
  Phys.\ Rev.\ D {\bf 89}, no. 11, 116011 (2014),
  [arXiv:1404.5577 [hep-ph]].

\bibitem{Farias:2014eca} 
  R.~L.~S.~Farias, K.~P.~Gomes, G.~I.~Krein and M.~B.~Pinto,
  Phys.\ Rev.\ C {\bf 90}, no. 2, 025203 (2014),
  [arXiv:1404.3931 [hep-ph]].

  
\bibitem{Ferreira:2014exa} 
  M.~Ferreira, P.~Costa, and C.~Providência,
  Phys.\ Rev.\ D {\bf 90}, no. 1, 016012 (2014),
  [arXiv:1406.3608 [hep-ph]].
 
\bibitem{Ayala:2014iba} 
  A.~Ayala, M.~Loewe, A.~J.~Mizher, and R.~Zamora,
  Phys.\ Rev.\ D {\bf 90}, no. 3, 036001 (2014),
  [arXiv:1406.3885 [hep-ph]].
  
\bibitem{Ayala:2014gwa} 
  A.~Ayala, M.~Loewe, and R.~Zamora,
  Phys.\ Rev.\ D {\bf 91}, no. 1, 016002 (2015),
  [arXiv:1406.7408 [hep-ph]].

\bibitem{Ayala:2015lta} 
  A.~Ayala, C.~A.~Dominguez, L.~A.~Hernandez, M.~Loewe, and R.~Zamora,
  Phys.\ Rev.\ D {\bf 92}, no. 9, 096011 (2015),
  Addendum: [Phys.\ Rev.\ D {\bf 92}, no. 11, 119905 (2015)],
  [arXiv:1509.03345 [hep-ph]].
  
\bibitem{Ayala:2015bgv} 
  A.~Ayala, C.~A.~Dominguez, L.~A.~Hernandez, M.~Loewe, and R.~Zamora,
  Phys.\ Lett.\ B {\bf 759}, 99 (2016),
  [arXiv:1510.09134 [hep-ph]].

\bibitem{Ayala:2016bbi} 
  A.~Ayala, C.~A.~Dominguez, L.~A.~Hernandez, M.~Loewe, A.~Raya, J.~C.~Rojas, and C.~Villavicencio,
  Phys.\ Rev.\ D {\bf 94}, no. 5, 054019 (2016),
  [arXiv:1603.00833 [hep-ph]].

\bibitem{Pagura:2016pwr} 
  V.~P.~Pagura, D.~Gomez Dumm, S.~Noguera, and N.~N.~Scoccola,
  Phys.\ Rev.\ D {\bf 95}, no. 3, 034013 (2017),
  [arXiv:1609.02025 [hep-ph]].

\bibitem{Costa:2013zca} 
  P.~Costa, M.~Ferreira, H.~Hansen, D.~P.~Menezes, and C.~Providência,
  Phys.\ Rev.\ D {\bf 89}, no. 5, 056013 (2014),
  [arXiv:1307.7894 [hep-ph]].

\bibitem{Avancini:2012ee} 
  S.~S.~Avancini, D.~P.~Menezes, M.~B.~Pinto, and C.~Providencia,
  Phys.\ Rev.\ D {\bf 85}, 091901 (2012),
  [arXiv:1202.5641 [hep-ph]].

\bibitem{Ruggieri:2014bqa} 
  M.~Ruggieri, L.~Oliva, P.~Castorina, R.~Gatto, and V.~Greco,
  Phys.\ Lett.\ B {\bf 734}, 255 (2014),
  [arXiv:1402.0737 [hep-ph]].

\bibitem{Costa:2015bza} 
  P.~Costa, M.~Ferreira, D.~P.~Menezes, J.~Moreira and C.~Providência,
  Phys.\ Rev.\ D {\bf 92}, no. 3, 036012 (2015),
  [arXiv:1508.07870 [hep-ph]].

\bibitem{Hatsuda:1994pi}
	T.~Hatsuda and T.~Kunihiro,
  Phys.\ Rep.\  {\bf 247}, 221 (1994),
  [hep-ph/9401310].

\bibitem{Klevansky:1992qe}
	S.~P.~Klevansky,
  Rev.\ Mod.\ Phys.\  {\bf 64}, 649 (1992).

\bibitem{Roessner:2006xn}
  S.~Roessner, C.~Ratti, and W.~Weise,
  Phys.\ Rev.\ D {\bf 75}, 034007 (2007),
  [hep-ph/0609281].

\bibitem{Rehberg:1995kh} 
  P.~Rehberg, S.~P.~Klevansky, and J.~Hufner,
  Phys.\ Rev.\ C {\bf 53}, 410 (1996),
  [hep-ph/9506436].
 
\bibitem{Menezes:2008qt.Menezes:2009uc} 
  D.~P.~Menezes, M.~B. Pinto, S.~S.~Avancini, A.~Perez Martinez, and C.~Provid\^{e}ncia,
  Phys.\ Rev.\ C {\bf 79}, 035807 (2009),
  [arXiv:0811.3361 [nucl-th]];
  D.~P.~Menezes, M.~B. Pinto, S.~S.~Avancini, and C.~Provid\^{e}ncia,
  Phys.\ Rev.\ C {\bf 80}, 065805 (2009),
  [arXiv:0907.2607 [nucl-th]].

\bibitem{Ebert:1999ht} 
  D.~Ebert, K.~G.~Klimenko, M.~A.~Vdovichenko and A.~S.~Vshivtsev,
  Phys.\ Rev.\ D {\bf 61}, 025005 (2000),
  [hep-ph/9905253].

\bibitem{Allen:2013lda} 
  P.~G.~Allen and N.~N.~Scoccola,
  Phys.\ Rev.\ D {\bf 88}, 094005 (2013),
  [arXiv:1309.2258 [hep-ph]].

\bibitem{Denke:2013gha} 
  R.~Z.~Denke and M.~B.~Pinto,
  Phys.\ Rev.\ D {\bf 88}, no. 5, 056008 (2013),
  [arXiv:1306.6246 [hep-ph]].

\bibitem{Grunfeld:2014qfa} 
  A.~G.~Grunfeld, D.~P.~Menezes, M.~B.~Pinto, and N.~N.~Scoccola,
  Phys.\ Rev.\ D {\bf 90}, no. 4, 044024 (2014),
  [arXiv:1402.4731 [hep-ph]].

\bibitem{Costa:2016vbb} 
  P.~Costa,
  Phys.\ Rev.\ D {\bf 93}, no. 11, 114035 (2016),
  doi:10.1103/PhysRevD.93.114035
  [arXiv:1610.06433 [nucl-th]].
  
\end{thebibliography}
\end{document}